# Variation of ionic conductivity in a plastic-crystalline mixture

D. Reuter, C. Geiß, P. Lunkenheimer[a)], and A. Loidl

*Experimental Physics V, Center for Electronic Correlations and Magnetism, University of Augsburg, 86135 Augsburg, Germany*

Ionically-conducting plastic crystals are possible candidates for solid-state electrolytes in energy-storage devices. Interestingly, the admixture of larger molecules to the most prominent molecular PC electrolyte, succinonitrile, was shown to drastically enhance its ionic conductivity. Therefore, binary mixtures seem to be a promising way to tune the conductivity of such solid-state electrolytes. However, to elucidate the general mechanisms of ionic charge transport in plastic crystals and the influence of mixing, a much broader data base is needed. In the present work, we investigate mixtures of two well-known plastic-crystalline systems, cyclohexanol and cyclooctanol, to which 1 mol% of Li ions were added. Applying differential scanning calorimetry and dielectric spectroscopy, we present a thorough investigation of the phase behavior and the ionic and dipolar dynamics of this system. All mixtures reveal plastic-crystalline phases with corresponding orientational glass-transitions. Moreover, their conductivity seems to be dominated by the "revolving-door" mechanism, implying a close coupling between the ionic translational and the molecular reorientational dynamics of the surrounding plastic-crystalline matrix. In contrast to succinonitrile-based mixtures, there is no strong variation of this coupling with the mixing ratio.

## I. INTRODUCTION

The extensive research on solid-state electrolytes for applications in various electrochemical devices is motivated by the shortcomings of their liquid counterparts. For example, in lithium-ion batteries solid-state electrolytes offer better packing efficiency, less costs, and more safety compared to the commonly used liquid electrolytes.[1,2] However, the big drawback of most solid electrolytes known today is their insufficient room-temperature ionic conductivity. Indeed, liquids are more suitable host systems for ions because their natural molecular diffusion enables high ionic mobility. Yet there are countless research efforts to overcome the problem of small room-temperature ionic conductivity in solid electrolytes.

The research on solid-state electrolytes covers various kinds of solid materials. For example, some crystalline materials like the so-called lithium superionic conductor $Li_{10}GeP_2S_{12}$ exhibit ionic conductivity up to 12 mS cm$^{-1}$ at room temperature.[3] Moreover, in the field of amorphous materials, polymers with admixed ions are extensively studied.[4,5] For the prominent polyethylene oxide it was shown that the ion transfer is strongly coupled to local segmental motions of the polymer chains, i.e. to the dynamics of the matrix.[6] However, the prime example of solids exhibiting local molecular motion are plastic crystals (PCs).[7] Consequently, PCs were found to reach technically relevant conductivity values at room temperature making them promising candidates for solid-state electrolytes.[8,9,10,11,12]

Plastic crystals naturally combine properties of liquid and solid matter. When transforming from the liquid to the PC state during cooling, the centers of mass of the molecules order on a regular crystalline lattice while the molecules retain their rotational degrees of freedom.[7] The molecules are translationally ordered but orientationally disordered and, therefore, this phase is also called "orientationally disordered crystal".[13] The transition is accompanied by a rather small entropy of fusion, $\Delta S_{fus} \leq 20$ J mol$^{-1}$ K$^{-1}$, which led to the discovery of the PC state as a distinct phase of solid materials.[14] They obtained their name from the high plasticity shown by many of these orientationally disordered phases, which is explained by the relatively weak interactions between the reorienting molecules.

The research on plastic-crystalline electrolytes can be divided into two groups. Ionic PCs that consist of at least two types of ionic molecules[15,16] and non-ionic PCs based on a single, neutral molecular species. The latter only exhibit technically relevant conductivity values when ions are added by admixing several mol% of a salt.[10,11,12] The most prominent of these materials is succinonitrile (SN), whose ionic conductivity for different ion species and concentrations was thoroughly investigated by Alarco et al.[11] More recently, the fundamental importance of the composition of the plastic-crystalline matrix for the ionic mobility was revealed: It was shown that mixtures of SN with the related molecular compound glutaronitrile (GN) have strongly enhanced ionic conductivity.[12,17,18] The mixtures exhibit stable PC phases between 15 and 70 mol% of GN[19,20] and their conductivity drastically increases with increasing amounts of GN.[12] This conductivity enhancement is accompanied by a successively stronger coupling of molecular reorientation and ionic translation. In Ref. 12 it was suggested that this behavior may arise from a "revolving door" or "paddle wheel" mechanism[9,16,21,22] that becomes optimized when replacing part of the smaller SN molecules in the crystalline lattice by the larger GN molecules. However, whether this is the sole reason for the observed conductivity enhancement remains unclear. Nonetheless, these results have opened up a new way to optimize PC electrolytes. Obviously, there is an urgent need for more data on the ionic conductivity in binary PC systems to better

---

[a)] Electronic mail: peter.lunkenheimer@physik.uni-augsburg.de

understand the physical reasons and mechanisms dominating the ionic mobility within the plastic-crystalline lattice.

In the present work, we report a thorough investigation of the phase behavior and the dielectric properties of binary mixtures of cyclohexanol (HEX) and cyclooctanol (OCT), two often investigated plastic-crystalline systems.[7,23,24,25,26] To introduce ions into these mixtures, we added 1% LiPF$_6$, a common salt for electrolytes which we also used in our previous investigation of the SN-GN system.[12] Just as for the latter, admixing OCT to HEX means a replacement of smaller by larger molecules. Moreover, in contrast to GN, which has no PC phase, for the HEX-OCT system both pure compounds form PC phases. They have relaxation times of $\tau_{HEX}(250\ K) = 1.4 \times 10^{-7}$ s (Ref. 7) and $\tau_{OCT}(250\ K) = 8.3 \times 10^{-7}$ s (Ref. 24), i.e., they exhibit significantly different local molecular dynamics. By mixing the two systems and adding 1 mol% of ions, we gradually vary the timescale of local molecular motion and, simultaneously, probe its influence on the ionic mobility. It should be noted that the choice of the materials for this work is not based on suitability for application but on the fact that HEX and OCT are widely studied plastic crystals, with similar shape but different size and significantly different relaxation times[7,25]. By providing information about another mixed system, in addition to SN-GN, we want to help achieving a better understanding of the principal mechanisms underlying the ionic conductivity in mixed plastic crystals.

## II. EXPERIMENTAL METHODS

HEX was purchased from Fluka (purity > 99 %) and OCT from Merck (purity > 95 %). LiPF$_6$ (purity 99.99 %) was obtained from Sigma-Aldrich. To enable a continuous study of the properties of the binary mixture HEX$_{1-x}$OCT$_x$, six concentrations between $x = 0$ and 1 were prepared. In addition, 1 mol% LiPF$_6$ was added to all mixtures. Calculated amounts of all components were mixed inside a glass tube, slightly heated ($T \approx 320$ K) and stirred on a magnetic stirrer for at least 12 h. In this way, complete solvation of the added salt was achieved for all samples.

Differential scanning calorimetry (DSC) measurements were performed using a DSC 8500 from Perkin Elmer. The samples were hermetically sealed into aluminum pans. The filled pans were placed in the sample chamber and an empty sealed pan was used as reference in the second chamber. The samples were cooled from 320 to 110 K and subsequently heated to 320 K applying a constant scanning rate of 10 K/min.

For the dielectric measurements, a frequency-response analyzer (Novocontrol Alpha-A analyzer) and an autobalance bridge (Agilent E4980A) were used. The liquid samples were filled into parallel-plate capacitors with plate distances of 0.1 mm. The temperature was regulated between 300 and 110 K with a N$_2$-gas flow cryostat.

## III. RESULTS AND DISCUSSION

All samples including the pure compounds were characterized using DSC with 10 K/min cooling/heating rates to check for phase- or glass-transition temperatures. The pure compounds were measured as received. Pure OCT forms a plastic-crystalline phase on cooling[26,27,28,29,30,31] revealed in the DSC trace by a slightly supercooled exothermic transition at about 280 K [Fig. 1(a), lower line]. Under sufficiently fast further cooling, a so-called glassy crystal with frozen orientational disorder is formed.[26,27,28,30] In Fig 1(a) it leads to a smeared-out anomaly at $T_g^o \approx 150$ K, whose shape resembles that expected for a glass transition. These results are in accord with the known sequence of transitions of cyclooctanol.[26,27,28,30] The detected glass transition marks the glass-like freezing of the reorientational motions of the molecules and is a well-known phenomenon for many PCs.[13,32,33,34] During heating [Fig. 1(a), upper line] the system returns from the glassy crystal into the PC phase at $T_g^o \approx 150$ K, again showing a glass-transition anomaly. This is followed by a sequence of cold crystallization and melting transitions between 240 K and 270 K that can be assigned to the formation and melting of phases with a higher degree of orientational order[26]. For OCT, different orientationally (or conformationally[35]) disordered and partially ordered phases are known[25,26,36,37]. The final melting of the fully orientationally disordered PC phase occurs at $T_m = 293$ K, in reasonable agreement with literature.[26,27,29,30]

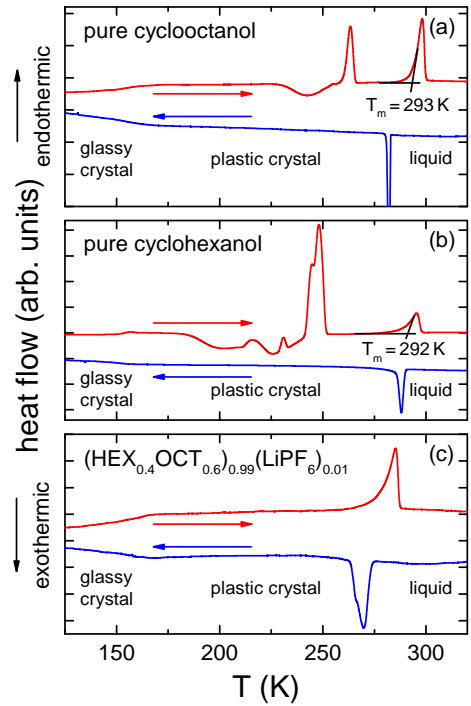

**FIG. 1.** DSC measurements on cooling (lower lines) and heating (upper lines) with rates of 10 K/min for pure cyclooctanol (a) and pure cyclohexanol (b), compared with the mixture (HEX$_{0.4}$OCT$_{0.6}$)$_{0.99}$(LiPF$_6$)$_{0.01}$ (c). Endothermic processes are plotted in positive $y$-direction. Three different phases are assigned to the cooling measurements as they sequentially occur on cooling: liquid, plastic crystal, and glassy crystal.

Pure HEX [Fig. 1(b)] displays a similar thermodynamic behavior with $T_g^o \approx 150$ K.[38,39] In this compound, the region of cold crystallization and subsequent melting during heating is spread from 180 K to 250 K, a signature of the complex phase diagram of cyclohexanol.[38,40] The fully orientationally disordered PC phase melts at $T_m$ = 292 K, which is consistent with literature.[23,38,40] Finally, as revealed by Fig. 1(c) (lower line) just as the pure compounds, the mixture $(HEX_{0.4}OCT_{0.6})_{0.99}(LiPF_6)_{0.01}$ also reveals a transition into the PC phase and a glass-like anomaly at $T_g^o$ under cooling. Obviously, glassy freezing of the orientational motions also occurs in this mixed system with added salt. Moreover, the absence of any additional anomalies in the heating measurement [Fig. 1(c), upper line] also implies homogeneity of the mixture and complete orientational disorder of its PC phase well up to the melting temperature.

exothermic transitions occurs between 180 K to 250 K, just as in pure HEX [Fig. 1(b)]. Furthermore, the salt addition reduces the melting temperatures $T_m$ of HEX and OCT roughly by 15 K compared to the pure compounds.

The fully orientationally disordered PC phase of HEX is known to have a face-centered cubic (fcc) structure[41,42,43]. On the other hand, the fully orientationally disordered PC phase of OCT was rather recently discovered to have a simple cubic (sc) lattice structure.[31] For a continuous mixing series of HEX with OCT, therefore, a transition from fcc to sc lattice structure has to emerge in the $T$-$x$ phase diagram. Such a transition could explain the unsystematic variation of melting temperature for $x \geq 0.8$. However, further investigations of the crystallographic structure of this mixed system are required to confirm this speculation.

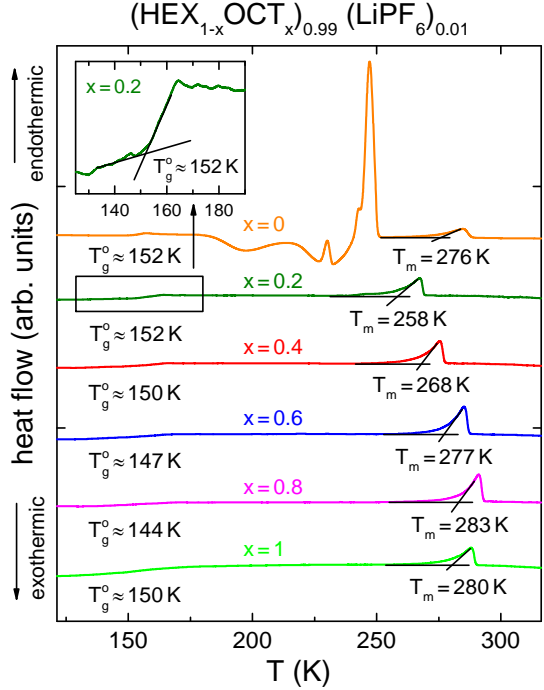

**FIG. 2.** DSC data taken on heating with a rate of 10 K/min for all mixtures, with endothermic processes pointing in positive y-direction. The data were recorded after cooling the samples from 320 to 110 K with rates of 10 K/min. The transition temperatures $T_g^o$ and $T_m$ (onset determination; solid lines) are indicated for each curve. The inset shows a magnification of the glass-transition step for $x = 0.2$.

In Fig. 2, the DSC measurements of all prepared mixtures containing LiPF$_6$ taken under heating are summarized and transition temperatures are indicated (the corresponding cooling curves qualitatively resemble those shown in Fig. 1). All samples exhibit an orientational glass-transition step close to $T_g^o \approx 150$ K (see inset for a zoomed view for $x = 0.2$) and a melting peak between 258 and 283 K. The orientational glass-transition and endothermic melting processes belong to the homogeneous PC phases of the mixtures as demonstrated in the discussion of Fig. 1. In contrast to pure OCT [Fig. 1(a)], no phases with a higher degree of orientational order are found for OCT ($x = 1$) with 1 mol% LiPF$_6$ during heating. Obviously, the small addition of 1 mol% salt suppresses their formation. In contrast, for HEX with 1 mol% salt ($x = 0$), a succession of endo- and

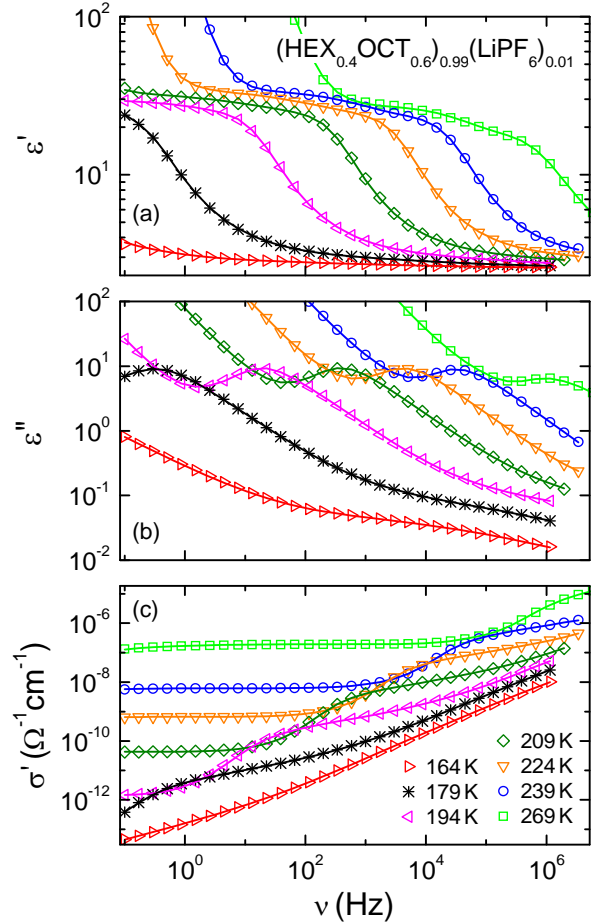

**FIG. 3.** Typical dielectric spectra of the PC phase of a HEX$_{1-x}$OCT$_x$ mixture with $x = 0.6$ and 1 mol% LiPF$_6$. The figure shows the dielectric constant (a), the dielectric loss (b), and the conductivity (c) for temperatures between 164 and 269 K. The solid lines are fits using an equivalent-circuit approach[44] consisting of a distributed RC circuit to account for electrode-polarization effects, a dc-conductivity contribution, and three relaxation functions (see text for details).

To obtain further insights into the molecular and ionic dynamics of these mixed molecular systems, dielectric spectroscopy was performed for all samples. The measurements were done during cooling from room temperature to 110 K. Figure 3 shows representative dielectric spectra for the mixture with $x = 0.6$ in the plastic-crystalline state. The dielectric constant $\varepsilon'$ [Fig. 3(a)] exhibits a step-like decrease from roughly 30 to 3 with increasing frequency. In accord with published results on the pure systems,[7,25] we ascribe this feature to the $\alpha$-relaxation process, i.e., the collective reorientational motions of the molecules sitting on the lattice sites of the PC. This notion is supported by the fact that the relaxational features found here for $x = 0$ and 1 have similar amplitude and frequency as previously reported for HEX and OCT without added salt.[7,25] The point-of-inflection frequency $\nu_p$ of the steps in Fig. 3(a) is related to the relaxation time of the $\alpha$ process via $\tau = 1/(2\pi\nu_p)$. Upon cooling, these steps systematically shift to lower frequencies indicating continuous slowing down of molecular motion, typical for glassy freezing.

For temperatures $T \geq 209$ K, a small second step seems to be superimposed on top of the main $\alpha$-relaxation step with amplitude $\Delta\varepsilon \approx 6$, which is most clearly seen at 269 K around about $10^5$ Hz. This process obviously occurs at lower frequencies and, thus, is slower than the $\alpha$ relaxation. A similar sub-$\alpha$ relaxation was also reported for pure cyclohexanol[23] and for the PC phase of SN-GN mixtures.[17,19] For SN-GN, it was suggested[17] to be due to space-charge effects due to spatial heterogeneities within the polycrystalline samples arising from a partial phase transformation below 233 K, where the plastic phase can transfer into the fully ordered crystal phase in pure SN. Generally, for heterogeneous samples additional apparent relaxation processes can appear, often termed Maxwell-Wagner relaxations.[44,45] As there is no anomaly in the dielectric properties of our mixtures that would point to such a phase transition, and as the additional slow relaxation process is observed up to the highest investigated temperatures, such an explanation, however, seems unlikely in the present case. An alternative explanation is an ac-conductivity contribution[19] as is common for hopping charge transport in disordered matter[46] or transitions between different conformers of the molecules.[19,23,47] Either way, the small sub-$\alpha$ relaxation does not notably influence the main relaxation and the dc conductivity of the samples and, therefore, a detailed analyses of this feature is out of the scope of the present work. Finally, $\varepsilon'$ for the three highest temperatures shown in Fig. 3(a) (224 to 269 K) exhibit an increase approaching colossal values in $\varepsilon'(\nu)$ at low frequencies due to extrinsic electrode-polarization effects.[44]

Corresponding to the steps of the main relaxation, in $\varepsilon''(\nu)$ [Fig. 3(b)] peaks are observed, e.g., for $T = 224$ K around 10 kHz. At the low-frequency flanks of these peaks, the dc-conductivity contribution leads to a strong increase of $\varepsilon''$ via the relation $\varepsilon'' \propto \sigma'/\nu$. The mentioned low-frequency relaxation also should lead to a peak in the loss, which, however, is strongly superimposed by this dc-conductivity contribution and not visible is this representation. For temperatures 164 to 194 K, at high frequencies a shoulder appears, indicating a secondary $\beta$-relaxation, as found for many PC systems, including HEX and OCT.[7,25] The conductivity $\sigma'$ shown in Fig. 3(c) essentially contains the same information as the dielectric loss [Fig. 3(b)]. However, this representation nicely reveals the frequency-independent region of $\sigma'(\nu)$ corresponding to the strongly temperature-dependent dc conductivity $\sigma_{dc}$ of the sample. It is best visible for $T = 269$ K between 1 Hz and 10 kHz where $\sigma_{dc}$ reaches values around $2\times10^{-7}$ $\Omega^{-1}$ cm$^{-1}$. In addition, for 269 K the conductivity decreases below about 1 Hz due to electrode-polarization effects. This effect is only visible for the highest temperature because, obviously, only there the ionic mobility is sufficiently high to lead to significant blocking of the electrodes before the field is reversed.[44]

We performed a detailed evaluation of the dielectric spectra using an equivalent-circuit approach, simultaneously fitting $\varepsilon'$ and $\varepsilon''$ (lines in Fig. 2).[44] The fit curves for $\sigma'$ were calculated via the relation $\varepsilon'' = \sigma'/(2\pi\nu\varepsilon_0)$ with $\varepsilon_0$ the permittivity of free space. To account for the electrode-polarization effects, a distributed RC circuit was used,[44,48] assumed to be connected in series to the bulk. The $\alpha$ relaxation was modelled by the empirical Cole-Davidson function[49], which usually provides a good description of the main relaxation in glass formers and PCs.[7,50,51] Two Cole-Cole functions[52] were employed to describe the $\beta$ and low-frequency relaxations.[7,53,54] In addition, a dc contribution $\varepsilon''_{dc} = \sigma_{dc}/(2\pi\nu\varepsilon_0)$ was used. (Depending on temperature, not all these elements were needed, e.g., at low temperatures, the electrode effects could be neglected.) Overall, the fits obtained in this way show perfect agreement with the experimental data. The relaxation-time and dc-conductivity values deduced from the fits match perfectly with the values directly read off from the spectra. Therefore, the following discussion relies on data directly determined from the spectra.

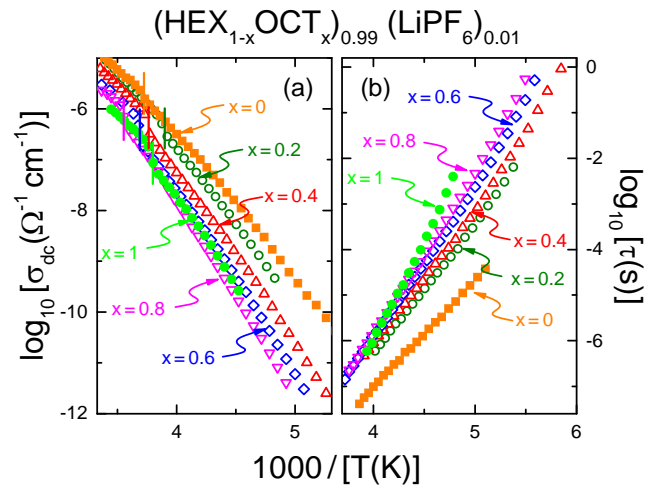

FIG. 4. Temperature-dependent dc conductivity (a) and reorientational $\alpha$- relaxation time (b) in the PC phases of all mixtures. Vertical lines in (a) indicate the melting temperatures. Slight deviations from those determined by DSC are due to supercooling effects.

The main focus of the present work is the determination of the relative changes of ionic conductivity in the binary PC system and its correlation with the molecular reorientational dynamics. In Fig. 4(a), the dc conductivity of all mixtures is plotted versus the inverse temperature in an Arrhenius representation. The melting temperatures are indicated by vertical lines. The conductivity only moderately drops when entering the PC phases, indicating high ionic mobility that to a large extend is favored by the reorientational molecular motion, present above and below $T_m$. Within the PC phases, there is a clear overall trend of a significant enhancement in conductivity (roughly 1.5 decades) when the *smaller* HEX molecules are admixed to the larger OCT molecules (i.e. when $x$ decreases from 1 to 0). This finding is in marked contrast to those for the SN-GN system, where the addition of the *larger* GN molecules leads to a conductivity increase up to three decades.[12] A closer inspection of Fig. 4(a) reveals that the addition of only 20 mol% HEX to OCT (i.e. going from $x = 1$ to $x = 0.8$) in fact first lowers $\sigma_{dc}$, before it shows the mentioned strong increase under further addition of HEX. Such discontinuity might well be explained by the transition from the sc to the fcc lattice structure around $x = 0.8$. Figure 5(a) shows the dependence of the dc conductivity on $x$ for 220 K, well within the PC phases. The conductivity continuously decreases with increasing $x$, however, with pure OCT not matching the general trend.

In Fig. 4(b) the reorientational relaxation times of the PC phases of all mixtures are presented in Arrhenius representation. $HEX_{0.99}(LiPF_6)_{0.01}$ ($x = 0$) reveals the fastest reorientational dynamics. While the relaxation times of the other mixtures nearly match at high temperatures when approaching the melting temperature, at lower temperatures they separate from each other. There, a continuous increase of $\tau$ with increasing amount of OCT is observed. Qualitatively similar behavior was found for the binary system cycloheptanol-cyclooctanol.[55] In Fig. 5(b) the concentration dependence of $\tau$ is shown for 220 K. It reveals a strong initial increase by about one decade when adding 20% OCT to HEX, followed by a continuous increase towards the relaxation time of $OCT_{0.99}(LiPF_6)_{0.01}$, which is about two decades slower than that of $HEX_{0.99}(LiPF_6)_{0.01}$.

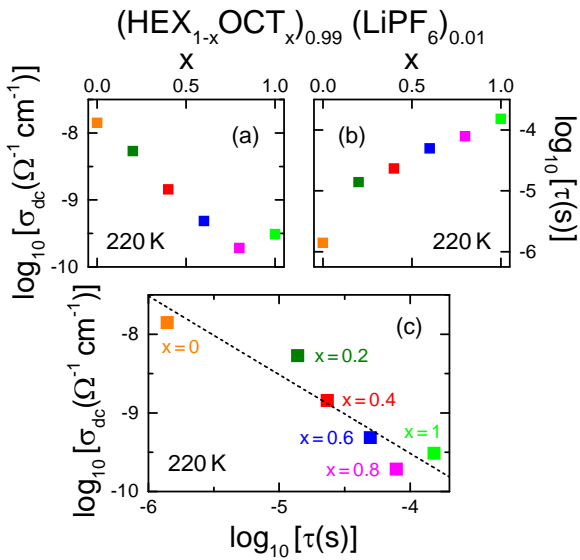

**FIG. 5.** (a) Mixing-ratio dependence of the dc conductivity and (b) of the reorientational $\alpha$-relaxation time for 220 K. (c) Combined data of (a) and (b) showing the dependence of the conductivity on relaxation time. The dashed line is a linear fit with fixed slope of -1, corresponding to $\sigma_{dc} \propto 1/\tau$.

In the Arrhenius plots of Fig. 4, most of the dc-conductivity and reorientational relaxation-time curves exhibit slight curvatures. This evidences small but significant deviations from thermally activated Arrhenius behavior. Such a non-Arrhenius temperature dependence of the conductivity is typical for ionic conduction in glasslike materials if there is some coupling of the ionic mobility to the glassy $\alpha$-relaxation dynamics.[6,12,18,56,57] Non-Arrhenius behavior of the latter is a hallmark feature of glassforming liquids[50,51,58] but is also found for PCs. However, there the deviations usually are weaker,[7,19,56,59] in accord with the relatively moderate deviations from linear behavior in Fig. 3.

To check for this coupling of ionic motion and $\alpha$-relaxation dynamics, a direct comparison of the temperature dependences of the reorientational relaxation times and the dc resistivity ($\rho_{dc} = 1/\sigma_{dc}$) provides further insights. In Fig. 6, for each mixture $\rho_{dc}(T)$ (squares; left axis) and $\tau(T)$ (circles; right axis) are shown in a common frame using an Arrhenius representation. For pure HEX with 1 mol% salt, below the melting point in the fully orientationally disordered PC phase (labeled PC I, Fig. 5), $\tau(T)$ can be fitted by an Arrhenius law [$\tau \propto \exp(E/T)$, with $E$ an energy barrier in K], in agreement with earlier findings.[7] On cooling, the transition into a phase with presumably increased orientational order was slightly supercooled to $T = 195$ K, where $\tau(T)$ shows a strong anomaly indicating the transition into a second PC phase (PC II).[7,26,40] However, in the PC phases of the HEX-OCT mixtures and in pure OCT, simple Arrhenius behavior is not sufficient to fit the experimental data. There $\tau(T)$ instead can be well described by the empirical Vogel-Fulcher-Tammann (VFT) law (solid lines through the circles in Fig. 6), commonly used for canonical glass formers and PCs.[7,19,25,50,51,53,58] Here the modified form[60]

$$\tau = \tau_0 \exp\left[\frac{DT_{VF}}{T - T_{VF}}\right] \qquad (1)$$

was used, where $\tau_0$ is a prefactor, $T_{VF}$ the Vogel-Fulcher temperature, and $D$ the so-called strength parameter. The latter is a measure of the so-called strength or fragility of a glass-forming system. Within the strong/fragile classification scheme of glass forming materials, low values of $D$ correspond to significant deviations from Arrhenius behavior (termed "fragile") while high values characterize so-called "strong" behavior with only weak deviations.[60,61] The obtained $D$ values, varying between 32 and 79, evidence relatively strong temperature characteristics of $\tau(T)$, in agreement with the findings for most PCs.[7,56]

The solid lines through the squares in Fig. 6 indicate fits of the dc resistivity. Similar to $\tau(T)$, to describe the temperature dependence of the dc resistivity for $x = 0$, an Arrhenius law was sufficient. However, $\rho_{dc}(T)$ of the mixtures and for $x = 1$ could be better fitted by a VFT law [eq. (1) with $\tau$ replaced by $\rho$]. For conventional ionic conductors, thermally activated Arrhenius behavior of $\rho_{dc}(T)$ would be expected. As mentioned above, the observed VFT behavior indicates glasslike freezing of the ionic mobility due to a (at least partial) coupling to the glassy reorientational dynamics of the PC molecules. That this coupling is close but not complete is revealed by the not quite perfect scaling of the $\rho_{dc}(T)$ and $\tau(T)$ curves in Fig. 6. This scaling seems to work somewhat better for the mixtures than for the pure systems but, overall, it varies only weakly with $x$. Interestingly, this finding again markedly contrasts with that in the SN-GN mixtures, where the scaling was found to clearly fail for pure SN but to become successively better when the larger GN molecules are added.[12]

By plotting the dc-conductivity versus the relaxation times for different $x$ in Fig. 5(c) for $T = 220$ K, the overall influence of the timescale of the molecular dynamics on the ionic translational motion can be studied. With $x$ increasing from zero to one, both the ionic dc conductivity and the molecular relaxation time vary by about two decades. This general trend indicates that the ionic mobility in this binary PC system is strongly supported by the molecular reorientational motion of the surrounding matrix. However, obviously the data do not follow the simple relation $\sigma_{dc} \propto 1/\tau$, which is indicated by the dashed line in Fig. 5(c). Especially, the deviation from the general decrease of $\sigma_{dc}(\tau)$, observed for $OCT_{0.99}(LiPF_6)_{0.01}$ ($x = 1$), may evidence a different ion-transport mechanism in pure OCT and/or reflect the mentioned transition from the fcc to the sc crystallographic structure when approaching high concentrations of OCT.

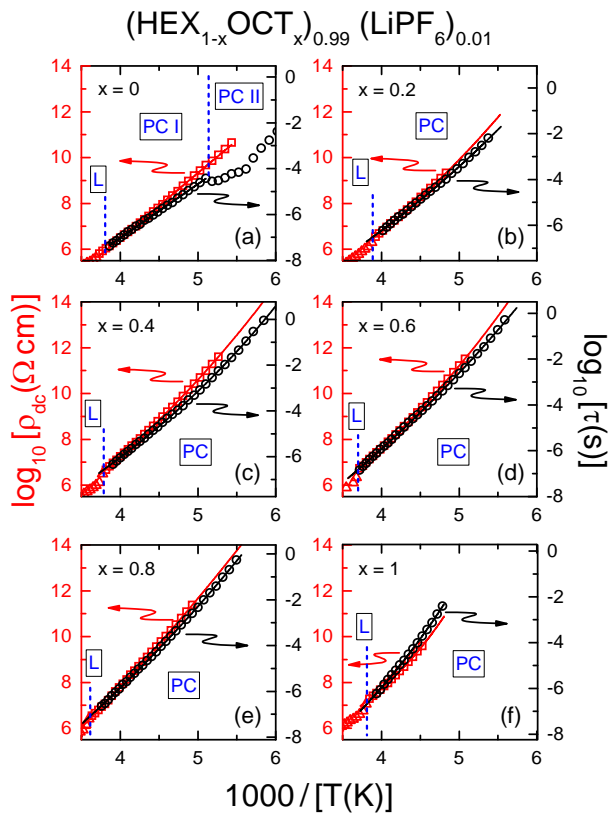

**FIG. 6.** Ionic and reorientational dynamics of all $HEX_{1-x}OCT_x$ mixtures with 1 mol% $LiPF_6$ plotted versus the inverse temperature (Arrhenius representation). The axis ranges of the dc-resistivity (squares; left axes) and relaxation-time data (circles; right axes) were chosen to cover the same number of decades and, in addition, were vertically shifted to match at the highest temperature in the PC phases. The vertical dashed lines indicate phase-transition temperatures, namely the liquid-PC transition (high temperatures) and, for low temperatures, the formation of a phase suggested to have higher orientational order,[26] named PC II (only found for $x = 0$). The solid lines are VFT fits of the data in the PC phases.

A correlation of ion diffusion and molecular reorientational dynamics is often regarded as evidence for the so-called "revolving door" or "paddle wheel" scenario[9,12,21,22] in ionically conducting PC systems. This picture highlights the importance of the on-site reorientational motion of the lattice molecules for the ionic mobility rather than vacancies and impurities[9,15,16,62,63,64,65] in the PC lattice. In the SN-GN system, the ionic conductivity was found to increase for higher GN contents[12] due to a successively better coupling of ionic and molecular motion. This was explained by a more effective revolving-door mechanism caused by the addition of the larger GN molecules. In contrast to the present HEX-OCT mixtures, the reorientational relaxation times of the SN-GN mixtures only weakly vary with the mixing ratio.[12] On the other hand, in the HEX-OCT system the ionic-molecular coupling does not seem to strongly vary (Fig. 6) and here it is the change of the relaxation time with $x$ that governs the conductivity variation. This explains the decrease of ionic conductivity when larger molecules are admixed, in contrast to the increase found for SN-GN. However, the deviations from strict inverse proportionality of $\sigma_{dc}$ and $\tau$, evidenced in Fig. 5(c), may well indicate a limited influence of other mechanisms. Either a variation of coupling as invoked for the SN-GN system[12], a defect-related conductivity as discussed, e.g., in Refs. 9,15,16,62,63,64,65 or the differences in lattice structures of pure HEX and OCT might explain the deviations from this simple inverse proportionality.

Overall, one has to be cautious when comparing absolute values of ionic conductivity of the SN-GN system with those of the HEX-OCT system: They might be strongly influenced by different interactions of the ions with the organic molecules, e.g., due to the hydrogen bonds present in the alcohol systems. Instead, the present work focuses on the principle mechanism of ionic conductivity in these systems thereby concentrating on the dependence of the ionic conductivity on molecular dynamics and on the coupling of these two dynamic processes in different molecular mixtures.

## IV. SUMMARY AND CONCLUSIONS

In the present work, we have provided a thorough investigation of the reorientational and ionic dynamics of the mixed PC system HEX-OCT with 1 mol% $LiPF_6$. While in the previously investigated SN-GN mixtures[12] an increasing content of the *larger* molecules causes a three-decade enhancement of the conductivity, for HEX-OCT we find that higher concentrations of the *smaller* molecules lead to an increase up to about 1-2 decades. This can be explained assuming that in HEX-OCT mixtures, via a revolving door mechanism, the ionic motion is always rather closely coupled to the reorientational one (which becomes faster when the smaller molecules are added). In contrast, in SN-GN this coupling strongly varies with the mixing ratio while the reorientational dynamics remains nearly unaltered.[12]

One may speculate that this qualitatively different behavior is related to the roughly disc-shaped form of the HEX and OCT molecules, in contrast to the three-dimensional, bulkier shapes of the SN and GN molecules, which in addition exist in rather different molecular conformations. In PCs, at least at high temperatures, isotropic reorientational motions can be assumed, leading to a time-averaged spherical shape of the molecules and the typical cubic crystal structures of these materials. For the disc-shaped HEX and OCT molecules, for short time intervals parallel orientations of neighboring molecules can be imagined that correspond to widely opened gaps, through which the ions can easily pass, i.e. the revolving-door scenario is highly effective and dominates the charge transport in these PCs. For the bulkier SN this obviously is less the case and, following the arguments in Ref. 12, only the addition of the larger GN molecules leads to effective revolving doors. While these considerations certainly seem plausible, we cannot prove their validity based on the present dielectric results alone and more investigations on further mixed PC systems and also using different experimental methods like nuclear magnetic resonance and x-ray diffraction are necessary.

Our results demonstrate that mixing different compounds showing plastic crystalline phases has considerable impact on the technically relevant ionic conductivity of PCs. However, they also reveal that this impact can be governed by different mechanisms, depending on the chosen plastic-crystalline materials. Obviously, the timescale of the reorientational processes in the PC phase can be an essential factor for the ionic mobility.

## ACKNOWLEDGEMENTS

This work was supported by the Deutsche Forschungsgemeinschaft (grant No. LU 656/3-1).

**References**

[1] J. M. Tarascon, and M. Armand, Nature **414**, 359 (2001).
[2] J. Motavalli, Nature **526**, 96 (2015).
[3] N. Kamaya, K. Homma, Y. Yamakawa, M. Hirayama, R. Kanno, M. Yonemura, T. Kamiyama, Y. Kato, S. Hama, K. Kawamoto, and A. Mitsui, Nat. Mater. **10**, 682 (2011).
[4] F. Croce, G. B. Appetecchi, L. Persi, and B. Scrosati, Nature **394**, 456 (1998).
[5] A. M. Stephan, and K. S. Nahm, Polymer **47**, 5952 (2006).
[6] C. Do, P. Lunkenheimer, D. Diddens, M. Götz, M. Weiß, A. Loidl, X.-G. Sun, J. Allgaier, and M. Ohl, Phys. Rev. Lett. **111**, 018301 (2013).
[7] R. Brand, P. Lunkenheimer, and A. Loidl, J. Chem. Phys. **116**, 10386 (2002).
[8] D. R. MacFarlane, J. H. Huang, and M. Forsyth, Nature **402**, 792 (1999).
[9] D. R. MacFarlane, and M. Forsyth, Adv. Mater. **13**, 957 (2001).
[10] S. Long, D. R. MacFarlane, and M. Forsyth, Solid State Ionics **161**, 105 (2003).
[11] P. J. Alarco, Y. Abu-Lebdeh, A. Abouimrane, and M. Armand, Nat. Mater. **3**, 476 (2004).
[12] K. Geirhos, P. Lunkenheimer, M. Michl, D. Reuter, and A. Loidl, J. Chem. Phys. **143**, 081101 (2015).
[13] E. Tombari, and G. P. Johari, J. Chem. Phys. **142**, 104501 (2015).
[14] J. Timmermans, J. Phys. Chem. Solids **18**, 1 (1961).
[15] J. M. Pringle, P. C. Howlett, D. R. MacFarlane, and M. Forsyth, J. Mater. Chem. **20**, 2056 (2010).
[16] J. M. Pringle, Phys. Chem. Chem. Phys. **15**, 1339 (2013).
[17] M. Zachariah, M. Romanini, P. Tripathi, M. Barrio, J. L. Tamarit, and R. Macovez, J. Phys. Chem. C **119**, 27298 (2015).
[18] M. Zachariah, M. Romanini, P. Tripathi, J. Ll. Tamarit, and R. Macovez, Phys. Chem. Chem. Phys. **17**, 16053 (2015).
[19] T. Bauer, M. Köhler, P. Lunkenheimer, A. Loidl, and C. A. Angell, J. Chem. Phys. **133**, 144509 (2010).
[20] M. Götz, T. Bauer, P. Lunkenheimer, and A. Loidl, J. Chem. Phys. **140**, 094504 (2014).
[21] R. Aronsson, B. Jansson, H. Knape, A. Lundén, L. Nilsson, C.-A. Sjöblom, and L. Torell, J. Phys. Colloq. **41**, 35 (1980).
[22] E. I. Cooper, and C. A. Angell, Solid State Ionics **18-19**, 570 (1986).
[23] L. P. Singh, and S. S. N. Murthy, Phys. Chem. Chem. Phys. **11**, 5110 (2009).
[24] Gangasharan, and S. S. N. Murthy, J. Chem. Phys. **99**, 9865 (1993).
[25] R. Brand, P. Lunkenheimer, and A. Loidl, Phys. Rev. B **56**, R5713 (1997).
[26] D. Leslie-Pelecky, and N. Birge, Phys. Rev. B **50**, 13250 (1994).
[27] A. Dworkin, A. H. Fuchs, M. Ghelfenstein, and H. Szwarc, J. Physique-Lettres **43**, L 21 (1982).
[28] M. Shablakh, L. Dissado, and R. Hill, J. Chem. Soc. Faraday Trans. 2, **79**, 369 (1983).
[29] R. Edelmann, and A. Wurflinger, Mol. Cryst. Liq. Cryst. **148**, 249 (1987).
[30] M. Tyagi and S. S. N. Murthy, J. Chem. Phys. **114**, 3640 (2001).
[31] M. A. Rute, J. Salud, P. Negrier, D. O. López, J. Ll. Tamarit, R. Puertas, M. Barrio, and D. Mondieig, J. Phys. Chem. **107**, 5914 (2003).
[32] O. Yamamuro, H. Yamasaki, Y. Madokoro, I. Tsukushi, and T. Matsuo, J. Phys.: Condens. Matter **15**, 5439 (2003).
[33] O. Yamamuro, M. Hayashi, T. Matsuo, and P. Lunkenheimer, J. Chem. Phys. **119**, 4775 (2003).
[34] H. Suga, J. Phys.: Condens. Matter **15**, S775 (2003).
[35] R. Puertas, M. A. Rute, J. Salud, D. O. López, S. Diez, J. Kees van Miltenburg, L. C. Pardo, J. Ll. Tamarit, M. Barrio, M. A. Pérez-Jubindo, and M. R. de la Fuente, Phys. Rev. B **69**, 224202 (2004).
[36] O. Andersson, and R. G. Ross, Mol. Phys. **71**, 523 (1990).
[37] H. Forsman, and O. Andersson, J. Non-Cryst. Solids **131-133**, 1145 (1991).
[38] K. Adachi, H. Suga, and S. Seki, Bull. Chem. Soc. Jpn. **41**, 1073 (1968).
[39] M. Mizukami, H. Fujimori, and M. Oguni, Solid State Commun. **100**, 83 (1996).
[40] K. Adachi, H. Suga, S. Seki, S. Kubota, S. Yamaguchi, O. Yano, and Y. Wada, Mol. Cryst. Liq. Cryst. **18**, 345 (1972).
[41] D. André, D. Ceccaldi, and H. Szwarc, J. Phys. Fr. **45**, 731 (1984).
[42] D. Ceccaldi, Phys. Rev. B **31**, 8221 (1985).
[43] B. Ibberson, S. Parsons, D. R. Allan, and A. M. T. Bell, Acta Cryst. B **64**, 573 (2008).
[44] S. Emmert, M. Wolf, R. Gulich, S. Krohns, P. Lunkenheimer, and A. Loidl, Eur. Phys. J. B **83**, 157 (2011).
[45] P. Lunkenheimer, S. Krohns, S. Riegg, S. G. Ebbinghaus, A. Reller, and A. Loidl, Eur. Phys. J. Special Topics **180**, 61 (2010).
[46] A. K. Jonscher, *Dielectric Relaxations in Solids* (Chelsea Dielectrics, London, 1983).
[47] M. Bée, R. E. Lechner, J. P. Amoureux, and R. Fouret, J. Phys. C **16**, 4973 (1983).
[48] B. Renner, P. Lunkenheimer, M. Schetter, A. Loidl, A. Reller, and S. G. Ebbinghaus, J. Appl. Phys. **96**, 4400 (2004).
[49] D. W. Davidson and R. H. Cole, J. Chem. Phys. **19**, 1484 (1951).
[50] P. Lunkenheimer, U. Schneider, R. Brand, and A. Loidl, Contemp. Phys. **41**, 15 (2000).
[51] P. Lunkenheimer, and A. Loidl, Chem. Phys. **284**, 205 (2002).
[52] K. S. Cole and R. H. Cole, J. Chem. Phys. **9**, 341 (1941).
[53] A. Kudlik, S. Benkhof, T. Blochowicz, C. Tschirwitz, and E. A. Rössler, J. Mol. Structure **479**, 201 (1999).
[54] S. Kastner, M. Köhler, Y. Goncharov, P. Lunkenheimer, and A. Loidl, J. Non-Cryst. Solids **357**, 510 (2011).

[55] J. C. Martínez-García, J. Ll. Tamarit, L. C. Pardo, M. Barrio, S. J. Rzoska, and A. Droz-Rzoska. J. Phys. Chem. B **114**, 6099 (2010).

[56] F. Mizuno, J.-P. Belieres, N. Kuwata, A. Pradel, M. Ribes, and C. A. Angell, J. Non-Cryst. Solids **352**, 5147 (2006).

[57] P. Sippel, P. Lunkenheimer, S. Krohns, E. Thoms, and A. Loidl, Sci. Rep. **5**, 13922 (2015).

[58] J. C. Dyre, Rev. Mod. Phys. **78**, 953 (2006).

[59] L. C. Pardo, P. Lunkenheimer, and A. Loidl, J. Chem. Phys. **124**, 124911 (2006).

[60] C.A. Angell, "Strong and fragile liquids," in *Relaxations in Complex Systems*, edited by K. L. Nai and G.B. Wright (NRL, Washington, DC, 1985), pp. 3-11.

[61] C. A. Angell, J. Non-Cryst. Solids **102**, 205 (1988).

[62] Y. Abu-Lebdeh, P.-J. Alarco, and M. Armand, Angew. Chem. Intl. Edn. **42**, 4499 (2003).

[63] J. Huang, A. Hill, M. Forsyth, D. R. MacFarlane, and A. Hollenkamp, Solid State Ionics **177**, 2569 (2006).

[64] J. Adebahr, A. J. Seeber, D. R. MacFarlane, and M. Forsyth, J. Phys. Chem. B **109**, 20087 (2005).

[65] D. Hwang, D. Y. Kim, S. M. Jo, V. Armel, D. R. MacFarlane, D. Kim, and S.-Y. Jang, Sci. Rep. **3**, 3520 (2013).